\documentclass[twocolumn]{aastex631}

\usepackage{amsmath} % or simply amstext
\usepackage{rotating} % <-- HERE
\usepackage{booktabs}

\newcommand\mm{\mathrm{mm}}

\newcommand\mJy{\mathrm{mJy}}

\newcommand\GHz{\mathrm{GHz}}
\newcommand\Jykms{\mathrm{Jy\,km\,s^{-1}}}
\newcommand{\angstrom}{\textup{\AA}}

\defcitealias{Edenhofer+2023}{E23}
\defcitealias{Leike+2020}{L20}

\received{February 02, 2024}
\revised{ April 05, 2024  }
\accepted{ April 06, 2024 }

\graphicspath{{./}{plots/}}

\begin{document}

\title{High-resolution Pan-STARRS and SMA observations of IRAS 23077+6707: A giant edge-on protoplanetary disk}

\author[0000-0002-5688-6790]{Kristina Monsch}
\affiliation{Center for Astrophysics $\vert$ Harvard \& Smithsonian, 60 Garden Street, Cambridge, MA 02138, USA}

\author[0000-0002-4248-5443]{Joshua Bennett Lovell}
\affiliation{Center for Astrophysics $\vert$ Harvard \& Smithsonian, 60 Garden Street, Cambridge, MA 02138, USA}

\author[0000-0001-5538-5903]{Ciprian T. Berghea}
\affiliation{U.S. Naval Observatory (USNO), 3450 Massachusetts Avenue NW, Washington, DC 20392, USA}

\author[0000-0003-3122-4894]{Gordian Edenhofer}
\affiliation{Center for Astrophysics $\vert$ Harvard \& Smithsonian, 60 Garden Street, Cambridge, MA 02138, USA}
\affiliation{Max Planck Institute for Astrophysics, Karl-Schwarzschild-Str. 1, 85741 Garching, Germany}
\affiliation{Ludwig Maximilian University of Munich, Geschwister-Scholl-Platz 1, 80539 Munich, Germany}

\author[0000-0002-3490-146X]{Garrett K. Keating}
\affiliation{Center for Astrophysics $\vert$ Harvard \& Smithsonian, 60 Garden Street, Cambridge, MA 02138, USA}

\author[0000-0003-2253-2270]{Sean M. Andrews}
\affiliation{Center for Astrophysics $\vert$ Harvard \& Smithsonian, 60 Garden Street, Cambridge, MA 02138, USA}

\author{Ammar Bayyari}
\affiliation{Department of Physics and Astronomy, University of Hawaii, Honolulu, HI 96822, USA}

\author[0000-0002-0210-2276]{Jeremy J. Drake}
\affiliation{Center for Astrophysics $\vert$ Harvard \& Smithsonian, 60 Garden Street, Cambridge, MA 02138, USA}
\affiliation{Lockheed Martin, 3251 Hanover St, Palo Alto, CA 94304}

\author[0000-0003-1526-7587]{David J. Wilner}
\affiliation{Center for Astrophysics $\vert$ Harvard \& Smithsonian, 60 Garden Street, Cambridge, MA 02138, USA}

\email{kristina.monsch@cfa.harvard.edu}

\begin{abstract}
We present resolved images of IRAS~23077+6707 (``Dracula's Chivito'') in $1.3\,\mm$/$225\,\GHz$ thermal dust and CO gas emission with the Submillimeter Array (SMA) and optical ($0.5$--$0.8\,\micron$) scattered light with the Panoramic Survey Telescope and Rapid Response System (Pan-STARRS).
The Pan-STARRS data show a bipolar distribution of optically scattering dust that is characteristic for disks observed at high inclinations. Its scattered light emission spans ${\sim}14''$, with two highly asymmetric filaments extending along the upper bounds of each nebula by ${\sim}9''$.
The SMA data measure $1.3\,\mm$ continuum dust as well as $^{12}$CO, $^{13}$CO and C$^{18}$O $J$=2$-$1 line emission over $12''$--$14''$ extents, with the gas presenting the typical morphology of a disk in Keplerian rotation, in both position-velocity space and in each CO line spectrum.
IRAS~23077+6707 has no reported distance estimate, but if it is located in the Cepheus star-forming region (180--800\,pc), it would have a radius spanning thousands of au.
Taken together, we infer IRAS~23077+6707 to be a giant and gas-rich edge-on protoplanetary disk, which to our knowledge is the largest in extent so far discovered.
\end{abstract}

%\keywords{Protoplanetary disks (1300) -- Young stellar objects (1834) -- Herbig Ae/Be stars (723)}

%TC:endignore
\section{Introduction} 
\label{sec:introduction}

Protoplanetary disks provide the physical and thermo-chemical resources for planet formation, and are a natural, ubiquitous by-product of the star formation process \citep[see][for reviews]{WilliamsCieza2011, Andrews+2020}. 
The vast majority have been discovered in surveys of relatively nearby star-forming regions \citep[see e.g. the review of][]{Manara+2023}, and specifically in star-forming regions such as Taurus \citep{Andrews+2013}, Ophiuchus \citep[][]{Cieza+2019}, Lupus \citep[][]{Ansdell+2016, Ansdell+2018, Lovell+2021} and Upper Scorpius \citep[][]{Carpenter+2014, Barenfeld+2016}, enabling the detailed structural mapping of disk morphologies and substructures \citep[e.g.][]{Andrews+2018, Long+2018}. Such detailed studies allow for connections to be made between these planetary birth-sites and the observed population of planets that are now readily detected around mature stars \citep[e.g.][]{Manara+2018, Mulders+2021}.

Measuring the internal structures of protoplanetary disks via different observational tracers is crucial for gaining a comprehensive understanding of protoplanetary disks and the intricate processes driving the birth of planetary systems.
Protoplanetary disks that are observed at high inclinations offer a unique opportunity to study their physical properties, as the bright emission of the central star is obscured entirely by the dust and gas of the surrounding disk, resulting in the disk being seen as a central dark lane flanked by scattered light emission \citep[e.g.][]{Burrows+1996, Padgett+1999, Wolf+2003, WatsonStapelfeldt2004, Stapelfeldt+2014}.
Consequently, these so-called `edge-on disks' (EODs) are special laboratories to study the environment in which planets are forming.
They allow us to directly observe the vertical disk structure that is needed in order to determine the degree of dust settling towards the midplane, but also to measure other key properties of planet-forming disks, such as their radii or their radial flaring. Comparison of scattered light images with extensive radiative transfer models further allows determining the physical properties of the scattering dust grains in the upper atmosphere of the disk.

The combination of scattered light imaging with interferometric observations in the millimeter (mm) regime has proven to be especially successful in characterizing EODs in detail \citep[e.g.][]{Wolf+2003, Bujarrabal+2008, Bujarrabal+2009, Sauter+2009, Wolff+2021}, as both types of observations trace different regions of a system and thus unique physical processes. 
On one hand, (sub-)mm observations are sensitive to the longer wavelength radiation being re-emitted from larger (mm- to cm-sized) dust grains close to the disk midplane, and thus typically appear in extended line (or `needle-like') morphologies.
On the other hand, optical and near-infrared (NIR) observations trace smaller, micron-sized dust grains in the hotter, outer envelope and the disk surface layers that scatter the light of the central star, and thus create two bright, highly-flared lobes.
Thanks to the \textit{Karl G. Jansky} Very Large Array (VLA), the Atacama Large Millimeter Array (ALMA), the \textit{Hubble} Space Telescope (HST), and most recently the \textit{James Webb} Space Telescope (JWST), scattered light and (sub-)mm observations of EODs at high spatial resolution have now become routine \citep[e.g.][]{Melis+2011, Stapelfeldt+2014, Villenave+2020, Villenave+2024, Flores+2021, Duchene+2023}. 
However, the angular extents of these EODs are generally only a few arcseconds at most, with IRAS~18059-3211 (``Gomez's Hamburger'') presenting the largest yet, with dust and gas extents of $9''$ and $12''$ respectively \citep{Ruiz+1987, Bujarrabal+2008, Bujarrabal+2009, Teague+2020}.

In this letter, we present the first resolved observations of IRAS~23077+6707 \citep[termed ``Dracula's Chivito", or ``DraChi" by][]{Berghea+2024} with the Submillimeter Array (SMA) and the Panoramic Survey Telescope and Rapid Response System (Pan-STARRS). 
IRAS~23077+6707 is located in the Cepheus star-forming region and was serendipitously discovered in 2016 by Dr. Ciprian Berghea in Pan-STARRS data.
Here we report IRAS~23077+6707 as the largest protoplanetary disk on the sky, with dust and CO gas extents of at least $12''$--$14''$, and CO in Keplerian rotation. 

The retrieval of the observational data and their reduction are described in \S\,\ref{sec:observations}. We present and discuss our results in \S\,\ref{sec:results}, and summarize our work in \S\,\ref{sec:conclusion}.

\section{Observations and analysis} 
\label{sec:observations}

\begin{figure*}[t!]
    \centering
    \includegraphics[width=\linewidth]{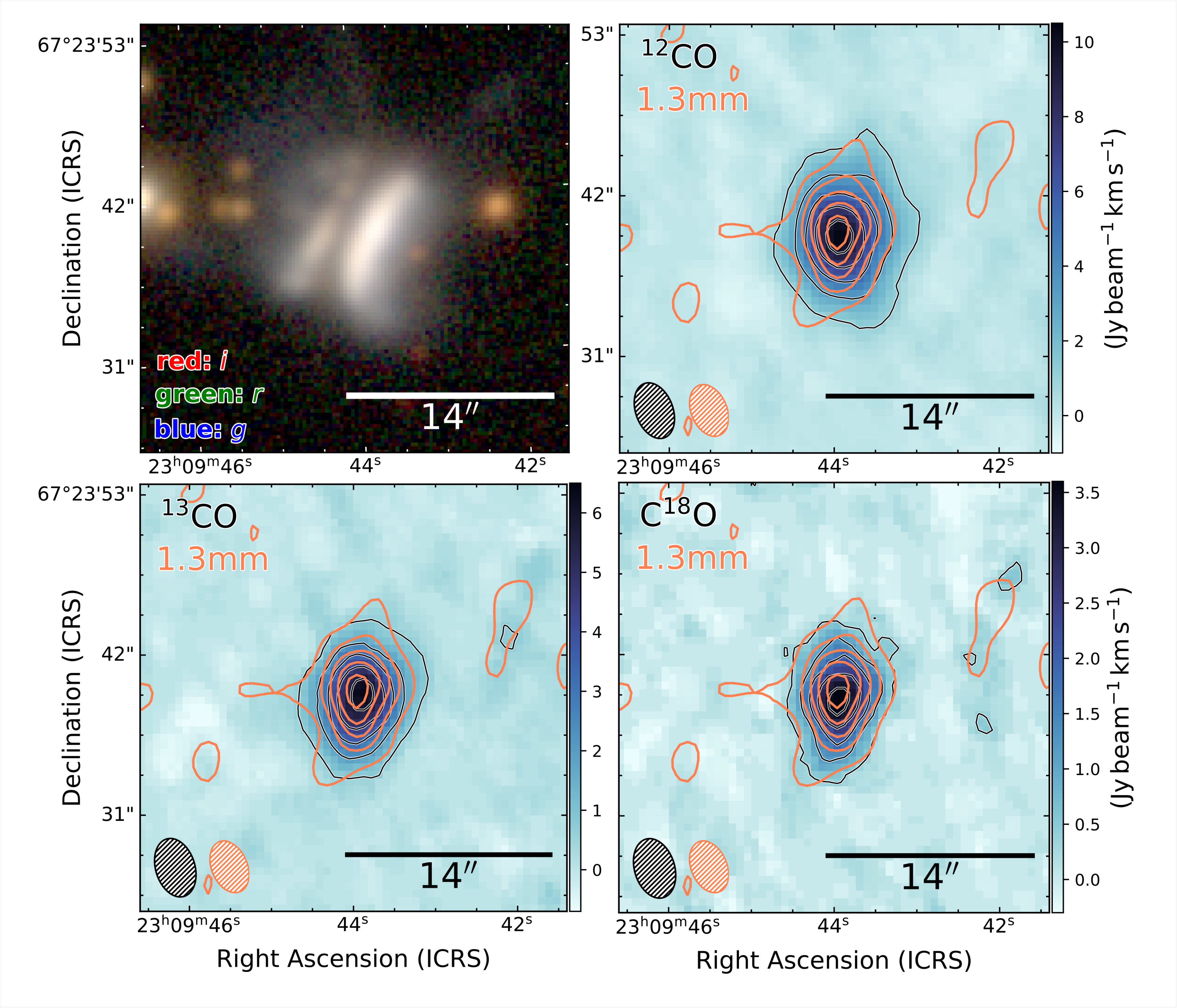}
    \caption{Comparison of optical scattered light and (sub-)mm emission of IRAS 23077+6707. \textit{Top left}: PS1 $irg$-composite image, showing the optical scattered light emission at $\lambda_\mathrm{mean}=7545$, 6215 and $4866\,\angstrom$, respectively. \textit{Top right}: SMA $^{12}$CO integrated intensity (or `moment 0') map, overlaid with its respective contours, as well as the 1.3\,mm continuum contours, drawn at the $10\,\%$, $30\,\%$, $50\,\%$, $70\,\%$ and $90\,\%$ levels of their corresponding maximum emission. \textit{Bottom left}: Same for $^{13}$CO. \textit{Bottom right}: Same for C$^{18}$O. In the lower right of each panel, we show $14''$ scale bars, and the effective CO (black) and continuum (orange) beams in the lower left of the SMA image.}
    \label{fig:mainFig}
\end{figure*}

\subsection{Panoramic Survey Telescope and Rapid Response System}
\label{sec:obs_pan-starrs}

Pan-STARRS \citep{PS1_Chambers+2016} is a system of two optical $1.8\,$m telescopes (PS1 and PS2) and the observations of IRAS~23077+6707 presented here were performed with the PS1 Gigapixel Camera (GPC1), covering a field of view of $\approx 7\,\mathrm{deg}^2$, a pixel scale of $0\farcs257$ and median seeing values of $1''$--$1\farcs3$ for the $grizy_\mathrm{P1}$ broadband filters (spanning $\lambda_\mathrm{mean}=4866$--$9633\,\angstrom$). 
Stacked images of IRAS~23077+6707 ($\mathrm{RA=23^{h}09^{m}43.645^{s}}$ and $\mathrm{Dec}=+67^\circ23'38\farcs94$), were retrieved from the first data release of PS1\footnote{The PS1 data are available at MAST: \dataset[doi:10.17909/s0zg-jx37]{\doi{10.17909/s0zg-jx37}}}.\citep{PS1_Flewelling+2020, PS1_DOI}. These were then used to produce the RGB-composite image shown in Fig.~\ref{fig:mainFig} (left) where the $i$-, $r$- and $g$-bands ($\lambda_\mathrm{mean}=7545$, $6215$ and $4866\,\angstrom$) are represented by red, green and blue colors respectively, and with equal image intensity weighting.

IRAS~23077+6707 shows the typical morphological features of an edge-on protoplanetary disk, namely two elongated bright reflection nebulae (tracing the upper layers of the disk atmosphere), separated by a dark lane (tracing the disk midplane) which obscures the light emitted by the central star.
The scattered light emission of the disk dark lane has an angular extent of $\approx 14''$ and a position angle of $338^\circ$ (from north, counter-clockwise). 
Further, two faint filaments that likely trace the arcs of the disk's flared upper layers, extend northwards by almost $9''$. 
We see no evidence of a jet nor any obvious signatures of envelope material, which are both typically associated with early-stage Class~0 and Class~I young stellar objects \citep[YSOs; see][for a review]{WilliamsCieza2011}, thus likely suggesting a more evolved evolutionary stage for IRAS~23077+6707.

IRAS~23077+6707 exhibits three prominent asymmetries that we derive in detail in Appendix~\ref{appendix:radial_intensity}.
First, its western lobe is brighter than its eastern lobe by a factor of 6, typical for disks with sky-projected inclinations that are not perfectly edge-on (due to preferential grain scattering), for which a simple visual comparison with the models of \citet[][see their Fig.\,4, in which the disks shown in the top row display brightness variations for different disk inclinations]{Watson+2007_PP5} implies the disk is likely inclined to the plane of the sky by $\approx80^\circ$. 
Second, in the eastern (fainter) lobe, the southern region is brighter than the northern one by a factor of 3, whereas in the western (brighter) lobe the northern and southern region brightnesses are within 1\,\%. 
Lateral asymmetries can be a result of foreground extinction preferentially dimming the north-east of the disk. However, they could also imply a misaligned, inner disk or a warp that would cast shadows on the observed disk (see Sect.~\ref{sec:results_DraChi_asymmetries}).

Third, whilst the northern extent has two prominent filaments, the southern side hosts none. These filaments extend northwards beyond the edge of the dark lane by $\approx 9''$.
The publicly available Digitized Sky Survey images of IRAS~23077+6707 also resolve the disk, and likewise demonstrate similar asymmetries (in DSS1, DSS2-blue, DSS2-red, and DSS2-infrared), which we thus deem a physical feature of the disk \citep{DSS1, Lasker1996A}.
The presence of such a filament-asymmetry could imply that the outermost disk layers (or possibly a disk wind) are truncated, or perhaps angled backwards, plausibly due to interactions with a nearby star and/or the ISM, or from asymmetric ionization/wind launching from the central star.

\subsection{Submillimeter Array}
\label{sec:obs_sma}

\subsubsection{Data reduction and calibration}
\label{sec:obs_sma_reduction}

%The Submillimeter Array (SMA) is an 8-dish (sub-)millimeter interferometer based on Maunakea, Hawai'i \citep{Ho2004}.
IRAS~23077+6707 was observed on 25 April 2023 by the SMA \citep{Ho+2004} in project 2022B-S054 (PI: K. Monsch), for 228\,min on-source.
The SMA was in the compact configuration with 6 antennas (1, 3, 4, 5, 6 and 8) spanning baselines of 6--67\,m.
The observations were conducted with the SMA SWARM correlator, consisting of $24\times2\,\GHz$ bands with 140\,kHz channel resolution tuned to a central local oscillator (LO) frequency of 225.538\,GHz ($\lambda=1.33\,\mathrm{mm}$), spanning 209.538--221.538\,GHz and 229.538--241.538\,GHz.
J0019+734 and J2005+778 were observed for gain calibration, 3c279 for bandpass calibration and Ceres for flux calibration.
We converted the raw SMA data to the \textit{Common Astronomy Software Applications} ({\tt CASA}) measurement set format with {\tt pyuvdata} \citep{Hazelton+2017}, using the SMA reduction pipeline in {\tt CASA} version 6.4.1\footnote{The SMA {\tt CASA} reduction pipeline can be accessed via: \url{https://github.com/Smithsonian/sma-data-reduction}} with a channel re-binning factor of 2 (to reduce the data volume, whilst maintaining high spectral resolution during calibration, appropriate for both continuum and spectral line analyses).
Prior to any calibration solutions being applied, we manually flagged all narrow interference spurs (that appeared in a small number of individual channels), and trimmed away 2.5\,\% of the overlapping ``guard-band'' regions between individual 2\,GHz SWARM correlator segments.

\subsubsection{Visibility modeling and imaging}
\label{sect:obs_sma_imaging}

`Corrected' data for IRAS~23077+6707 were extracted using the {\tt CASA mstransform} task, in which we time-averaged our data by $20\,\mathrm{s}$. 
For the continuum emission, we re-binned these corrected data to 4 channels per $2\,\GHz$ spectral window (appropriate for continuum analysis), whereas for line emission analysis, we split these corrected data into the 12 separate SMA spectral windows (with no channel averaging) and transformed these to the local standard of rest frame (LSRK).

We measure the total 1.3\,mm flux of IRAS~23077+6707 as $F_{\rm{1.3\,\mm}}=44.0\pm2.5$\,mJy by fitting a Gaussian model to IRAS~23077+6707's calibrated visibilities with the {\tt CASA} package {\tt uvmodelfit} (with 10 iterations to ensure model convergence, where the error incorporates a standard SMA 5\,\% flux calibration error in quadrature with the fitting uncertainty).
This fit derives a position angle of 342$^\circ$ and a small phase offset ($-0\farcs2$ in RA, and $0\farcs65$ in Dec).
By instead fitting the same visibility model to the upper and lower side bands of the continuum visibilities independently (between 209.5--221.5\,GHz and 229.5--241.5\,GHz) we found best-fit fluxes of $37.2\pm3.4$\,mJy and $52.6\pm3.8$\,mJy. This results in a spectral index of $\alpha=3.9\pm1.3$ (for $\alpha = \partial \log{F_\nu}/ \partial \log{\nu}$), which is at the higher end of mm spectral indices measured for other inclined disks \citep[e.g.][]{Ribas+2017, Villenave+2020}. Our value is consistent with either optically thin or optically thick emission, but due to the significantly shorter frequency range probed by our observations, our estimate of $\alpha$ carries substantial uncertainties and a direct comparison to these previous studies is not yet possible. Higher resolution observations with wider frequency baselines would be needed in order to discern between optically thin or optically thick dust in IRAS~23077+6707. 

Standard imaging of the continuum data was conducted with the {\tt tclean} algorithm to a $2\sigma$ threshold (${\sim}1.0$\,mJy), using an image-centered $16''\times8''$ elliptical mask (fully covering the disk), and a Briggs robust parameter of $0$.
The remaining panels of Fig.~\ref{fig:mainFig} show the resulting continuum emission overlaid as contours on the $^{12}$CO, $^{13}$CO and C$^{18}$O velocity-integrated intensity maps (or `moment 0 maps', which we will discuss further below), presenting an elongated morphology with a consistent position angle of ${\approx} 340^{\circ}$ with the scattered light emission, typical for an EOD imaged at relatively low-resolution where the $1.3\,\mm$ emission traces the cooler mid-plane dust. 
We deem the low axis ratio of the continuum extent (major/minor axis disk extent) to be dominated by the $3\farcs6\times2\farcs4$ beam roughly aligned along the disk minor axis with a BPA$=23^\circ$.

We subtract the continuum emission from each channel in our CO line measurement sets (for the spectral windows containing the $^{12}$CO, $^{13}$CO and C$^{18}$O) using the {\tt CASA} package {\tt uvcontsub} with a linear fit to all regions excluding the channels with CO emission ($\pm50\,$MHz).
We apply standard imaging on each CO line separately with the {\tt tclean} algorithm (to a clean threshold of $2\,\sigma \approx 0.1\,\mathrm{Jy}$) using an image-centered $16''\times8''$ elliptical mask, and a Briggs robust parameter of $0$) to produce 60-channel wide LSRK-frame data cubes, with the line emission spanning the central 10--20 channels. 
All cubes present spatially and spectrally resolved line emission at IRAS~23077+6707's location, with the resulting cubes having beams with sizes and position angles $3\farcs9\times2\farcs5$, BPA=$21^\circ$ ($^{12}$CO), $4\farcs0\times2\farcs6$, BPA=$17^\circ$ ($^{13}$CO), and $4\farcs1\times2\farcs7$, BPA=$17^\circ$ (C$^{18}$O).

\begin{figure*}[t!]
    \centering
    \includegraphics[width=0.95\linewidth]{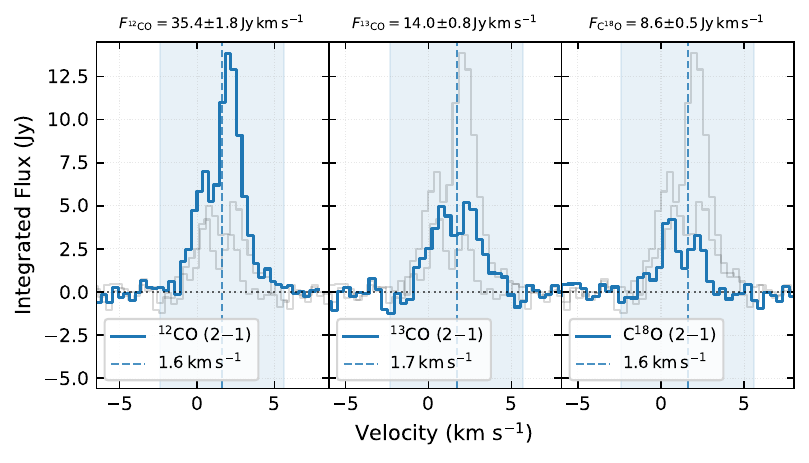}
    \caption{Line spectra of the $^{12}$CO, $^{13}$CO and C$^{18}$O $J$=2$-$1 lines. The line profiles were produced with \texttt{GoFish} \citep{GoFish} in a field of view spanning $22''$ centered on the respective line emission, where we adopted a disk inclination of $80^\circ$ and a position angle of $PA=342^\circ$. The resulting integrated line fluxes are $F_\mathrm{^{12}CO}=35.4\pm1.8\,\mathrm{Jy\,km\,s^{-1}}$, $F_\mathrm{^{13}CO}=14.0\pm0.8\,\mathrm{Jy\,km\,s^{-1}}$ and $F_\mathrm{C^{12}O}=8.6\pm0.5\,\mathrm{Jy\,km\,s^{-1}}$.} The light blue range highlights the velocity range in which the integrated intensity maps shown in Fig.~\ref{fig:mainFig} were computed.
    \label{fig:line_spectra}
\end{figure*}

In the top right and bottom panels of Fig.\,\ref{fig:mainFig}, we present the moment 0 maps for the CO emission lines, generated with \texttt{bettermoments} \citep{TeagueForeman-Mackey2018} using a $2\sigma$ emission clip that is integrated only from those channels of the data cube that span the full line width (which ranges from $-2.4$ to $5.6\,$km\,s$^{-1}$, cf. Fig.~\ref{fig:line_spectra}).
All three gas lines are clearly detected, with disk emission (detected at a $3\sigma$ level) covering more than $\sim14''$ for both $^{12}$CO and $^{13}$CO and $\sim 12''$ for C$^{18}$O (i.e. 3--4 beams across the radial disk extent). 
Notably, while being significantly less abundant than $^{12}$CO, the extents of both the $^{13}$CO and C$^{18}$O emission along the radial and vertical directions are comparable to the $^{12}$CO emission, revealing this as a highly gas-rich system. 

With {\tt GoFish} \citep{GoFish} and a simple $22''$ field of view cut centered on the respective line emission, we extract integrated line fluxes for the CO lines of $F_{^{12}\rm{CO}}=35.4\pm1.8\,\mathrm{Jy\,km\,s^{-1}}$, $F_{^{13}\rm{CO}}=14.0\pm0.8\,\mathrm{Jy\,km\,s^{-1}}$ and $F_{\rm{C}^{18}\rm{O}}=8.6\pm0.5\,\mathrm{Jy\,km\,s^{-1}}$ (which we report here and in Table~\ref{tab:summary} with an additional 5\% flux uncertainty in quadrature with the {\tt GoFish} fitting error). This results in flux ratios of $F_\mathrm{^{12}CO}/F_\mathrm{^{13}CO}\approx2.5$ and $F_\mathrm{^{13}CO}/F_\mathrm{C^{18}O}\approx1.6$. 
As expected for disks in Keplerian rotation, the line profiles for all three CO isotopologues as shown in Fig.~\ref{fig:line_spectra} are double-peaked, with velocity centroids of 1.6, 1.7 and 1.6\,km\,s$^{-1}$ respectively. Out of these, the $^{12}$CO line profile is the only observed as asymmetric, with its redshifted (western) component being $\approx 2\times$ brighter than its blueshifted (eastern) component, which is most likely a consequence of the $^{12}$CO emission being optically thick, preventing us from observing as much CO on the disk rear side.

\begin{figure*}[t!]
    \centering
    \includegraphics[width=\textwidth]{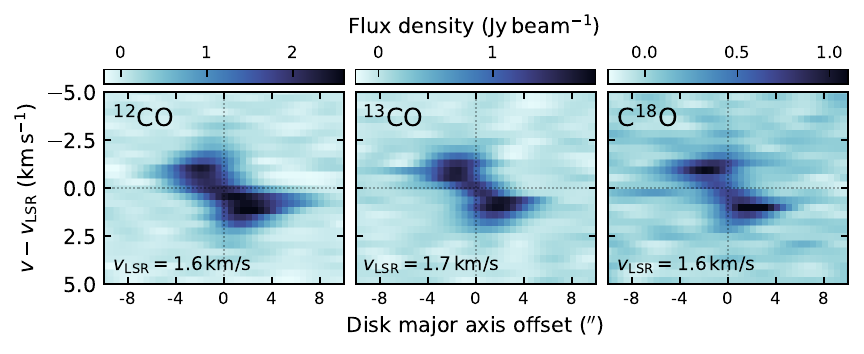}
    \caption{Position-velocity (PV) diagrams obtained for the three CO $J$=2$-$1 lines, with $22''$ long and $6''$ wide cuts, oriented along a position angle of $342^{\circ}$ from the image center, referenced to each of their respective line velocity centers.}
    \label{fig:pvdiagrams}
\end{figure*}

In Fig.~\ref{fig:pvdiagrams}, we further show position-velocity (PV) diagrams for all three CO $J$=2$-$1 lines, extracted with {\tt pvextractor} \citep{Ginsburg+2016} using $22''$ long and $6''$ wide cuts (which was chosen in order to encompass the $5\,\sigma$ contour of the $^{12}$CO emission along the disk minor axis), oriented along a position angle of $342^\circ$ and referenced to each of their respective line velocity centers, as well as the center of each cut.
All of these PV diagrams present the expected double-peaked (`dumb-bell'-like) emission morphology of an inclined disk in Keplerian rotation.

\section{Results and discussion} 
\label{sec:results}

\subsection{IRAS~23077+6707: a young, edge-on disk}
\label{sec:results_DraChi_location}

Our observations present compelling evidence that the giant bipolar reflection nebula, IRAS~23077+6707, is a protoplanetary disk that is viewed nearly edge-on.
Whilst its size is extreme, with IRAS~23077+6707 hosting a protoplanetary disk that subtends the largest known angular scale on the sky, what is perhaps most remarkable is that this source has until now gone unreported in the scientific literature.
We note that in the recent work of \citet{Angelo2023} it was reported that the number count of EODs falls far short of simulated disk populations in the Galaxy. The serendipitous discovery of IRAS~23077+6707 therefore suggests that there may be many more EODs awaiting discovery.

Nevertheless, decades of all-sky mapping, alongside detailed local Galaxy studies have reported unresolved detections of dust and gas associated with IRAS~23077+6707 \citep[for example, at moderate angular resolution with IRAM, WISE and 2MASS, see][]{Wouterloot1989, Wouterloot1993, Skrutskie+2006, Cutri+2014}.
Using these all-sky data, as well as employing the spectral slope expression pioneered by \citet{Lada1987} and \citet{, Adams1987}:

\begin{equation}
    \alpha_{\rm{IR}} = \frac{\partial \log{\nu F_\nu}}{\partial \log{\nu}},
\end{equation}
we determine IRAS~23077+6707 to have mid-IR spectral slopes of $\alpha_{\rm{2-12\mu m}}=-0.21$, and $\alpha_{\rm{2-22\mu m}}=0.1$, between the 2MASS $K$-band data point ($2.16\,\mu$m, $39.2{\pm}0.8$\,mJy) and either the WISE W3 ($12\,\mu$m, $310.0{\pm}3.1$\,mJy) or WISE W4 ($22\,\mu$m, $318.0{\pm}5.3$\,mJy) photometry.
While these values would ordinarily place a disk in the spectral slope range defining a class~F YSO \citep[i.e. an intermediate class~I--II source, cf.][]{Greene1994, WilliamsCieza2011}, IRAS~23077+6707's edge-on morphology may suggest it is instead more evolved due to the emission being highly self-extincted at these near- and mid-infrared wavelengths \citep[cf.][]{ChiangGoldreich1999, Robitaille+2006}. 
Indeed, since class~II YSOs viewed edge-on present SEDs more consistent with class~I and F sources, we deem IRAS~23077+6707's photometry as evidence for it instead being a class~II YSO, and thus host to a mature protoplanetary disk, which is consistent with the scattered light imaging that is clearly dominated by disk emission.

\subsection{IRAS~23077+6707: a giant, gas-rich disk in Cepheus}
\label{sec:results_DraChi_Rdisk}

\begin{figure*}[t!]
    \centering
    \includegraphics[width=0.95\linewidth]{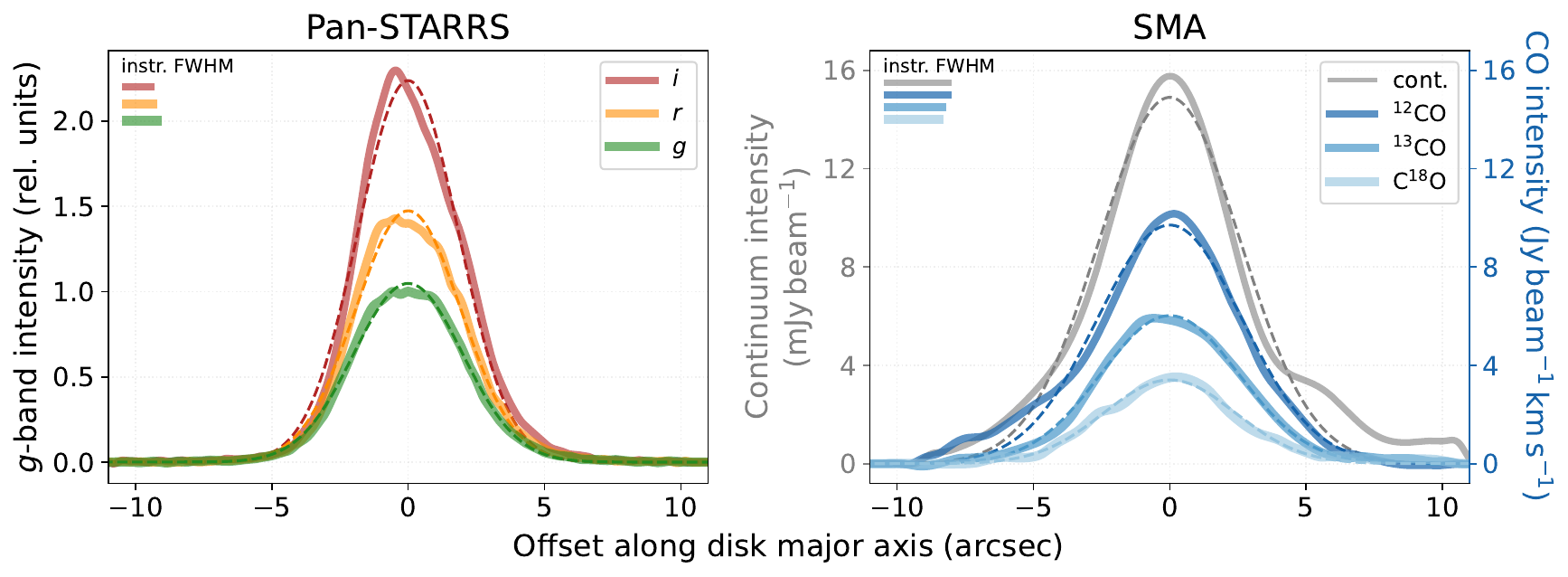}
    \caption{Radial intensity profiles of IRAS~23077+6707 computed along the disk major axis (midplane). \textit{Left:} PS1 $i$- (red), $r$- (orange) and $g$-band (green) data. Each line was normalized to the peak of the $g$-band profile. \textit{Right:} $1.3\,\mm$ continuum (gray), $^{12}$CO (dark blue), $^{13}$CO (mid blue), C$^{18}$O (light blue). By fitting 1D Gaussian models to each radial profile (with the instrumental FWHMs shown in the upper left), we estimate physical scales on which the disk emission can arise.}
    \label{fig:rad_profile}
\end{figure*}

Fig.~\ref{fig:rad_profile} shows the radial intensity profiles of the scattered light in the left panel (each normalized to the peak of the $g$-band intensity) and the $1.3\,\mm$ continuum, as well as the CO $J$=2$-$1 moment 0 maps in the right panel. Each profile reflects the emission of a single pixel lane along the disk major axis (i.e. midplane), either at a position angle of $338^{\circ}$ (PS1) or $342^{\circ}$ (SMA).\footnote{While the disk midplane appears ``dark'' in the PS1 images, the flux neither reaches zero nor noise level in these pixels and thus it is useful for measuring the disk's extent.} 
Whilst the first approximate radii at which the radial profiles of our data are consistent with noise 
%(i.e. within $1\sigma$ of $0\,\mathrm{mJy\,beam^{-1}}$ or $0\,\mathrm{Jy\,beam^{-1}\,km\,s^{-1}}$) 
are $8\farcs1$, $7\farcs7$, $7\farcs9$, $9\farcs1$, $8\farcs4$, $8\farcs7$, and $8\farcs5$ (for the $irg$ scattered
light, mm continuum, and progressively fainter CO lines respectively), convolution with either the PS1 PSF or SMA clean beams (with their respective FWHMs\footnote{The instrumental FWHMs (where $\mathrm{FWHM}=2\sqrt{2\ln2}\sigma$) are $1\farcs5$ (PS1 $g$), $1\farcs3$ (PS1 $r$), $1\farcs2$ \citep[PS1 $i$, cf.][]{PS1_Chambers+2016}, $2\farcs5$ ($1.3\,\mm$ continuum), $2\farcs5$ ($^{12}$CO), $2\farcs3$ ($^{13}$CO) and $2\farcs2$ (C$^{18}$O), respectively.} shown in the upper left corners of Fig.~\ref{fig:rad_profile}) suggests we are likely overestimating the disk's true angular extent. As such, we fit 1D Gaussian profiles using a least-squares method in order to provide an estimate for the scale on which dust and gas is plausibly present in the disk. These 1D Gaussian profiles offer a simple, but reasonable approximation of the disk emission, and thus allow us to estimate where the physical disk emission plausibly extends to, with these having deconvolved $2\,\sigma$ to $3\,\sigma$ radii of $3\farcs8$--$5\farcs7$, $3\farcs7$--$5\farcs6$, $3\farcs5$--$5\farcs3$, $4\farcs7$--$7\farcs1$, $4\farcs9$--$7\farcs3$, $4\farcs7$--$7\farcs0$ and $4\farcs5$--$6\farcs7$, respectively (see Table~\ref{tab:summary} for a summary of all measured values). We nevertheless note that the profiles, especially in the case of the scattered light, the continuum and the $^{12}$CO emission, host long tails that are not well-fitted by a simple Gaussian model, and consequently this Gaussian width approximation likely underestimates the disk's true physical size.

In summary, we find it highly plausible that the total extent of the disk's (sub-)mm emission is likely tracing dust and gas out to \textit{radii} of at least $\sim7''$, far larger than any other reported protoplanetary disks.
At the distance to the Cepheus star-forming region \citep[spanning a range of 180--800\,pc, see e.g.,][]{Kun+2008, Kun+2009, Szilagyi+2021, Kerr+2023} to which IRAS~23077+6707 is likely associated, such large radii would suggest that IRAS~23077+6707's disk thus spans 1000s of au in size.

What presents an interesting finding is that at the resolution of our observations, both IRAS~23077+6707’s dust continuum and gas emission span very similar spatial scales.
In other young planet-forming disks, the difference in gas-to-disk radii ratio is typically on the order of at least factor 2 or more \citep[cf.][]{BirnstielAndrews2014, Ansdell+2018, Facchini+2019, BoydenEisner2020, Trapman+2023}. While this suggests an immense reservoir of dust and gas present in IRAS~23077+6707, higher resolution data are needed to test if this disk indeed corresponds to an outlier among previously studied protoplanetary disks.

In addition, IRAS~23077+6707 is $4.3\times$ brighter in $^{12}$CO (2--1), $5.7\times$ brighter in $^{13}$CO (2--1) and over $9.4\times$ brighter in C$^{18}$O (2--1) than HD~34282 \citep[SpT A3, cf.][]{Merin+2004}, which is the brightest disk studied in a recent SMA survey of chemistry in Herbig Ae/Be disks (at similar resolution to our observations) performed by \citet{Pegues+2023}. In contrast, IRAS~23077+6707's $1.3\,\mm$ continuum emission is not only fainter than HD~34282 by a factor of $\sim3$, but also fainter than all other disks studied by \citet{Pegues+2023}. This may be a direct consequence of the different disk inclinations (e.g. $i\lesssim 60^\circ$ for the disks in \citealt{Pegues+2023}), and possibly the $1.3\,\mm$ continuum emission also being optically thick for IRAS~23077+6707. 

Only IRAS~23077+6707’s most well-known counterpart, Gomez's Hamburger \citep[SpT A0, cf.][]{Ruiz+1987}, measures a similarly high integrated intensity for the CO gas, with $37.2\,\Jykms$ for $^{12}$CO and $16.5\,\Jykms$ for $^{13}$CO \citep[][]{Teague+2020}, while having a significantly higher continuum flux of $293\,\mJy$ compared to the $44\,\mJy$ we measure from our observations (despite the similarly high inclination of $\sim 86^\circ$ for Gomez's Hamburger), which could indicate a higher gas-to-dust ratio in IRAS~23077+6707. 

In summary, at this point it remains unclear what is driving these differences in the dust and gas properties of IRAS~23077+6707, and observations at comparable resolution and sensitivity to these previous disk studies are needed for more direct comparisons to similar Herbig Ae/Be systems.

\subsection{IRAS~23077+6707: a misaligned system?}
\label{sec:results_DraChi_asymmetries}

The flux asymmetries that are observed in the PS1 image of IRAS~23077+6707, with the western lobe being brighter than its eastern one by a factor of 6, may be produced through different physical processes. 
Most likely, it is simply a geometric effect as a consequence of the disk not being observed perfectly edge-on, as previously discussed in Sect.~\ref{sec:obs_pan-starrs}. However, also perfectly edge-on systems have been found to display brightness asymmetries that may be possibly rooted in physical processes.

For instance, the formation of brightness asymmetries may happen in the earliest stages of a YSO during the late infall of gas after the initial collapse phase of the forming star \citep[e.g.][]{Kuffmeier+2021} or even during the class~II phase \citep[cf.][]{Ginski+2023}. 
Also, there is growing evidence that brightness asymmetries may have a planetary origin, either due to localized gravitational instabilities as a result of ongoing giant planet formation \citep[e.g.,][]{Morales+2019}, misaligned inner disks casting time-variable shadows on the outer disk due to warps created by super Jupiters \citep[e.g.][]{Nealon+2018, Debes+2023} or binary companions \citep{Facchini+2018}.

Indeed, in the PS1 data, IRAS~23077+6707 shows striking similarity with the simulated scattered light images presented by \citet[][see their Fig.~6]{Nealon+2019}. 
These authors found that when a planet is massive enough to carve a gap and separate the inner from the outer disk \citep[which is naturally the case for giant planets, cf.][]{LinPapaloizou1986a}, warps will form across the planetary orbit, resulting in a misalignment between the inner and outer disk \citep{Xiang-GruessPapaloizou2013, Nealon+2018}. 
Inner--outer disk misalignments of less than a degree are already able to cast shadows at larger stellocentric radii, resulting in brightness asymmetries that can be readily observed in scattered light emission in addition to the brightness asymmetry induced by the nearly edge-on orientations. 

With strong resemblance to the factor 3 brightness asymmetry along IRAS~23077+6707's eastern lobe, \citet{Villenave+2024} have very recently identified a lateral brightness symmetry (with a switch in brightness at $12.8$ and $21\,\micron$) in the edge-on Class~I system IRAS~04302+2247, which they attribute to a tilted inner disk. In addition, they uncover similar lateral asymmetries in optical and infrared scattered light emission in 15 out of 20 EODs, suggesting that a significant fraction of protoplanetary disks may be hosts to tilted inner regions. 
It is thus likely that also the presence of the lateral asymmetry in IRAS~23077+6707's eastern lobe may hint at the presence of a misaligned inner disk.

Regardless, whichever precise mechanism is driving the asymmetry in IRAS~23077+6707's disk, only higher-angular resolution scattered light observations in optical and especially near- to mid-infrared wavelengths \citep[e.g. in order to identify the switch in brightness due to an inner tilted disk, cf.][or the identification of an eccentric circumbinary disk producing similar brightness asymmetries, as e.g. for IRAS~04158+2805, cf. \citealt{Andrews+2008} \& \citealt{Ragusa+2021}]{Villenave+2024}, will be able to provide constraints on the possible origins of the asymmetries observed in IRAS~23077+6707.

\begin{table*}[ht!]
\centering
    \begin{tabular}{c|ccccccc}
    \hline \hline
     & \multicolumn{2}{c}{Flux} & $R_{2\sigma}$ & $R_{3\sigma}$ & Asymmetries \\
     & ($\mathrm{mJy}$) & ($\Jykms$) & ($''$) & ($''$) & \\
    \hline
    SL ($gri$) & -- & -- & (3.8,3.7,3.5) & (5.7,5.6,5.3) & 1) East-West brightness asymmetry ($\approx 6\times$) \\
    -- & -- & -- & -- & -- & 2) North-South brightness asymmetry in east lobe ($\approx 3\times$)  \\
    -- & -- & -- & -- & -- & 3) Two filaments extending northwards by $\approx 9''$  \\
    cont. & $44.0\pm2.5$ & -- & $4.7$ & $7.1$ & --\\
    $^{12}$CO & -- & $35.4\pm1.8$ & $4.9$ & $7.3$ & 4) West--East (blue--red shifted) line flux asymmetry ($\approx 2\times$) \\
    $^{13}$CO & -- & $14.0\pm0.8$ & $4.7$ & $7.0$ & -- \\
    C$^{18}$O & -- & $8.6\pm0.5$ & $4.5$ & $6.7$ & -- \\
    \hline
    
    \end{tabular}
    \caption{Summary of measured properties for IRAS~23077+6707. All reported millimeter fluxes include a 5\% flux uncertainty in quadrature with their respective fitting errors.}
    \label{tab:summary}
\end{table*}

\section{Summary and conclusions} 
\label{sec:conclusion}

In this work, we present the first resolved images of the giant edge-on protoplanetary disk IRAS 23077+6707 of thermal dust and gas emission with the Submillimeter Array (SMA) in direct comparison to optical scattered light with Pan-STARRS (PS1). 
Our main results can be summarized as follows:

\begin{enumerate}
    \item IRAS~23077+6707 shows a bright bipolar nebula, with a dark, optically obscured midplane indicative of a planet-forming disk that is observed nearly edge-on ($i\approx 80^\circ$).   
    
    \item IRAS~23077+6707's extent in optical scattered light emission spans $\sim14''$ on the sky, with a pair of asymmetric flared filaments additionally extending by $9''$ northwards. 

    \item The scattered light emission of the disk exhibits two further asymmetries. While the western lobe is brighter than its eastern lobe by a factor of 6 (which is typical for disks with sky-projected inclinations that are not perfectly edge-on), the eastern (fainter) lobe shows a north-south asymmetry, with the southern region being brighter by a factor of 3.
    
    \item Coincident with the scattered light disk traced by PS1, the SMA observations of IRAS~23077+6707 measure dust at $1.3\,\mm$ and $^{12}$CO, $^{13}$CO and C$^{18}$O $J$=2$-$1 line emission out to $12''$--$14''$ extents, with the gas being in Keplerian rotation.

\end{enumerate}

Overall, the discovery of IRAS~23077+6707 presents us with a golden opportunity to investigate planet formation in one of the largest planet-forming disks ever discovered.
The here presented observations provide the first joint optical and (sub-)mm analysis of this system.

%TC:ignore
\break
\textit{\large{Acknowledgments:  }}
%\begin{acknowledgments}
We thank the anonymous reviewer for their prompt and constructive report, which significantly helped to improve the clarity of this manuscript.
KM was supported by NASA {\it Chandra} grants GO8-19015X, TM9-20001X, GO7-18017X, and HST-GO-15326. 
JBL acknowledges the Smithsonian Institute for funding via a Submillimeter Array (SMA) Fellowship. 
GE acknowledges the support of the German Academic Scholarship Foundation in the form of a PhD scholarship (``Promotionsstipendium der Studienstiftung des Deutschen Volkes'').
JJD was funded by NASA contract NAS8-03060 to the {\it Chandra X-ray Center} and thanks the Director, Pat Slane, for continuing advice and support. 
The Submillimeter Array is a joint project between the Smithsonian Astrophysical Observatory and the Academia Sinica Institute of Astronomy and Astrophysics and is funded by the Smithsonian Institution and the Academia Sinica. The authors wish to recognize and acknowledge the very significant cultural role and reverence that the summit of Maunakea has always had within the indigenous Hawaiian community, where the Submillimeter Array (SMA) is located. We are most fortunate to have the opportunity to conduct observations from this mountain. We further acknowledge the operational staff and scientists involved in the collection of data presented here. The SMA data used here is from project 2022B-S054 and can be accessed via the Radio Telescope Data Center (RTDC) at \url{https://lweb.cfa.harvard.edu/cgi-bin/sma/smaarch.pl} after these have elapsed their proprietary access periods. 
The Pan-STARRS1 Surveys (PS1) and the PS1 public science archive have been made possible through contributions by the Institute for Astronomy, the University of Hawaii, the Pan-STARRS Project Office, the Max-Planck Society and its participating institutes, the Max Planck Institute for Astronomy, Heidelberg and the Max Planck Institute for Extraterrestrial Physics, Garching, The Johns Hopkins University, Durham University, the University of Edinburgh, the Queen's University Belfast, the Harvard-Smithsonian Center for Astrophysics, the Las Cumbres Observatory Global Telescope Network Incorporated, the National Central University of Taiwan, the Space Telescope Science Institute, the National Aeronautics and Space Administration under Grant No. NNX08AR22G issued through the Planetary Science Division of the NASA Science Mission Directorate, the National Science Foundation Grant No. AST-1238877, the University of Maryland, Eotvos Lorand University (ELTE), the Los Alamos National Laboratory, and the Gordon and Betty Moore Foundation. 
This publication makes use of data products from the Wide-field Infrared Survey Explorer, which is a joint project of the University of California, Los Angeles, and the Jet Propulsion Laboratory/California Institute of Technology, funded by the National Aeronautics and Space Administration. 
This publication makes use of data products from the Two Micron All Sky Survey, which is a joint project of the University of Massachusetts and the Infrared Processing and Analysis Center/California Institute of Technology, funded by the National Aeronautics and Space Administration and the National Science Foundation.
%\end{acknowledgments}


%\vspace{5mm}
\facilities{\textit{Submillimeter Array (SMA)}, \textit{Pan-STARRS}, \textit{WISE}, \textit{2MASS}}


\software{\texttt{APLpy} \citep{RobitailleBressert2012}, \texttt{Astropy} \citep{astropy2013, astropy2018, astropy2022}, \texttt{bettermoments} \citep[][]{TeagueForeman-Mackey2018}, \texttt{CASA} \citep{McMullin+2007}, \texttt{dustmaps} \citep{Green2018}, \texttt{GoFish} \citep{GoFish}, \texttt{Matplotlib} \citep{Hunter2007}, \texttt{NumPy} \citep{Harris+2020}, \texttt{pvextractor} \citep{Ginsburg+2016}, \texttt{pyuvdata} \citep{Hazelton+2017}}

\appendix
\section{Radial intensity profiles}
\label{appendix:radial_intensity}

\begin{figure}[t!]
    \centering
    \includegraphics[width=0.95\linewidth]{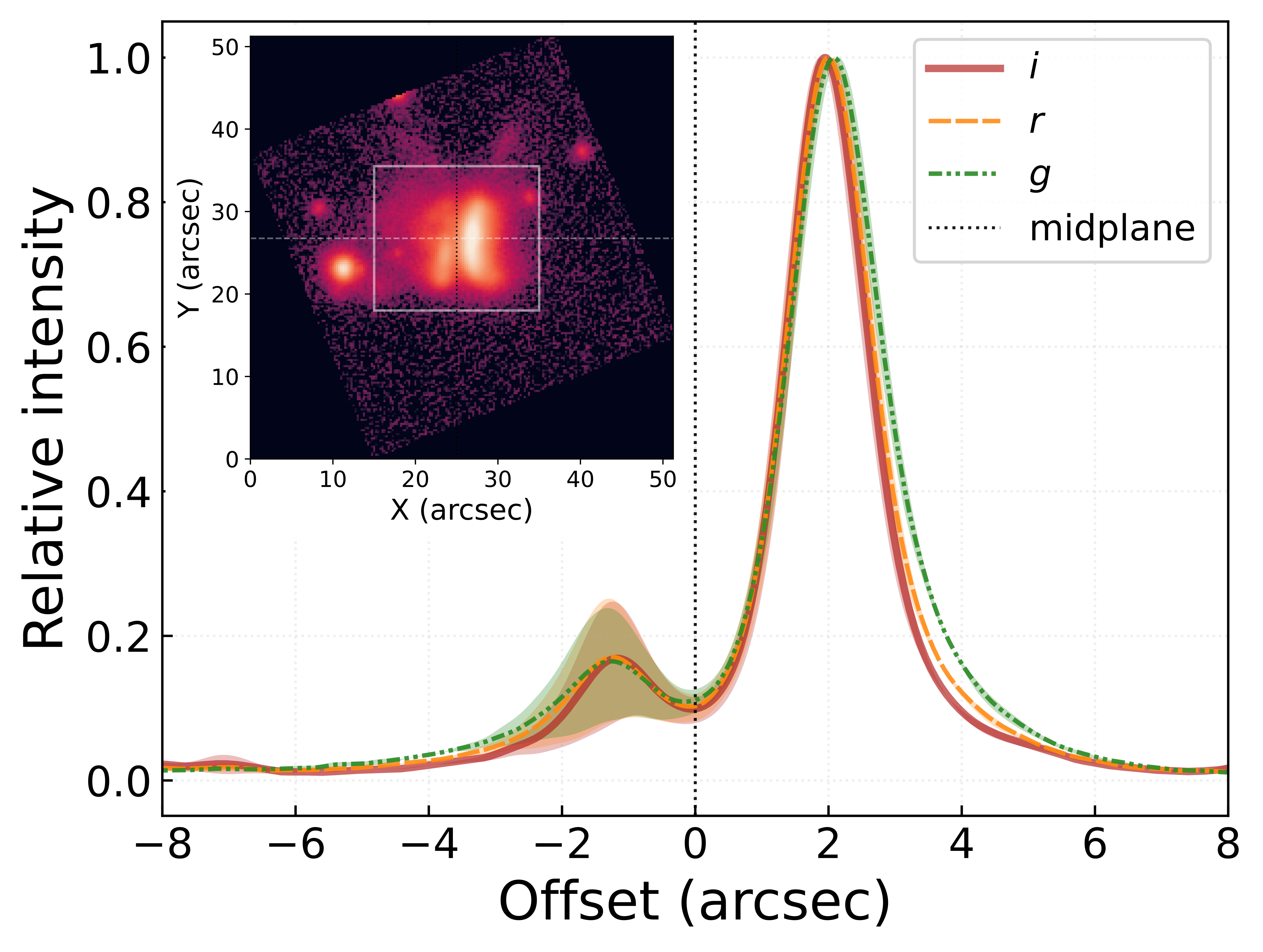}
    \caption{Calculation of IRAS~23077+6707's brightness asymmetries in scattered light. \textit{Left:} A $\sim 50''\times50''$ cutout (rotated by $22^\circ$ counter-clockwise) of the PS1 $g$-band filter, where the white box highlights the region in which the intensity profiles were computed. \textit{Right:} Emission profiles for the PS1 $i$- (solid-red), $r$ (dashed-green) and $g$- (dash-dotted blue line) filters. Each line corresponds to the mean intensity profiles of the full disk (i.e. averaged along all pixel lanes perpendicular to the disk midplane), that were additionally normalized to their corresponding maximum and subsequently centered on the disk's midplane (black-dotted line). The upper and lower bounds of each shaded region encompass the mean profiles of the southern and northern disk halves, respectively.  }
    \label{fig:rad_intensity}
\end{figure}

Figure~\ref{fig:rad_intensity} demonstrates our approach to measure IRAS~23077+6707's scattered light brightness asymmetries.
In the top left, we show a cutout (rotated by $22^\circ$ counter-clockwise) of the $g$-band filter, in which the white box highlights the region in which the intensity profiles were analyzed.
The right panel shows the mean emission profiles of the disk for each PS1 filter band that were computed perpendicular to the disk's midplane, each normed to their corresponding maximum and centered on the disk midplane.
A strong east-west brightness asymmetry becomes immediately apparent for all PS1 bands, with the western side of the disk being a factor $6\times$ brighter than the eastern one.

The \textit{upper and lower} bounds of the shaded regions shown for each line correspond to the mean profiles of the \textit{southern and northern} halves of the disk (divided by the white, dashed line in the $g$-band image cutout), respectively. These were obtained by averaging the emission along all Y-pixel lanes in each sub-cube. 
While less pronounced than the east-west asymmetry, a significant north-south symmetry along IRAS~23077+6707's eastern lobe can be measured, with the southern half being almost a factor $3\times$ brighter than the northern one. However, no prominent north-south asymmetry along the western lobe becomes apparent.

\bibliography{literature}{}
\bibliographystyle{aasjournal}


%TC:endignore
\end{document}